\documentstyle[12pt,epsfig]{article}

\font\twelverm=cmr10 scaled\magstep 1

\renewcommand\thefootnote{\,\arabic{footnote})}
\renewcommand\section[1]{%
\par\vspace{20pt}\noindent{\Large\bf #1\par}\vspace{14pt}}

\catcode`@ = 11
\renewcommand\@cite[2]{{#1\if@tempswa, #2\fi}}
\catcode`@ = 12

\renewenvironment{thebibliography}[1]{%
\newpage
\section{References}
\vspace{-\itemsep}
\vspace{-\parsep}
\baselineskip=12pt
\frenchspacing
\hyphenpenalty = 10000
\begin{list}{\arabic{enumi}.}
{\usecounter{enumi}
\setlength{\parsep}{0pt}
\setlength{\itemsep}{3pt}
\settowidth{\labelwidth}{#1.}}}
{\end{list}}

\newcommand\boldtau{%
\makebox[0.015em][l]{$\tau$}%
\makebox[0.015em][l]{$\tau$}%
\makebox[0.015em][l]{$\tau$}%
\makebox[0.015em][l]{$\tau$}%
\tau}

\newcommand\boldomega{%
\makebox[0.015em][l]{$\omega$}%
\makebox[0.015em][l]{$\omega$}%
\makebox[0.015em][l]{$\omega$}%
\makebox[0.015em][l]{$\omega$}%
\omega}

\begin{document}

\twelverm

\begin{center}
{\bf MASS SPLITTINGS OF NUCLEAR ISOTOPES IN CHIRAL SOLITON APPROACH\par}
\vglue 0.8cm
{\footnotesize
V.~B.~Kopeliovich\renewcommand\thefootnote{*}\footnote[0]{e-mail:
kopelio@al20.inr.troitsk.ru, kopelio@cpc.inr.ac.ru},
A.~M.~Shunderuk\renewcommand\thefootnote{**}\footnote[0]{e-mail:
A.Shunderuk@umail.ru},
G.~K.~Matushko\par}
{\footnotesize\em Institute for Nuclear Research, \\
the Russian Academy of Sciences, Moscow, Russia\par}
\vglue 0.1cm
\end{center}
\vglue 0.1cm
{\rightskip=3pc
 \leftskip=3pc
 \footnotesize
 \noindent
The differences of the masses of isotopes with atomic numbers between $\sim 10$
and $\sim 30$ can be described within the chiral soliton model in satisfactory
agreement with data. The rescaling of the model is necessary for this purpose ---
decrease of the Skyrme constant by $\sim 30 \%$, providing the ``nuclear variant''
of the model. The asymmetric term in Weizsacker---Bethe---Bacher mass
formula for nuclei can be obtained as the isospin dependent quantum
correction to the nucleus energy. Some predictions of the binding energies of
neutron rich isotopes are made in this way from, e.g. $^{16}{\rm Be}$,
$^{19}{\rm B}$ to $^{31}{\rm Ne}$ or $^{32}{\rm Na}$.
The neutron rich nuclides with high values of
isospin are unstable relative to the decay due to strong interactions.
The SK4 (Skyrme) variant of the model, as well as SK6 variant
(sixth order term in chiral derivatives in the Lagrangian as solitons 
stabilizer)
 are considered, the rational map approximation is used to 
describe multiskyrmions.
\vglue 0.3cm}
\vglue 0.05cm
\section{1. Introduction}
In the absence of the complete theory of strong interactions and nuclear
forces the checking of fundamental principles, which are believed to hold in 
any
 theoretical model, can be useful and of great importance. The description of
some well established and rather general properties of nuclei, as well as
searches and studies of unusual forms of matter, in particular the neutron rich
nuclides, may provide important source of lacking information and impact for
the development of new concepts and ideas.

The effective field theories are the powerful tool for the studies not
only of mesons, baryons and their interactions at low
energy [\cite{1}$-$\cite{3}], but also baryonic systems (nuclei) which appear as
quantized bound states of skyrmions with baryon number $B \geq 2$. The
properties of
the deuteron and $^3{\rm He}$, $^3{\rm H}$ and $^4{\rm He}$ were
explained semiquantitatively
[\cite{4}$-$\cite{5}] starting from few basic principles and ingredients contained
in the effective chiral Lagrangian of the Skyrme model [\cite{1}] or its
modifications (short review of early results can be found, e.g. in [\cite{6}]).

For the $B=2$ system the quantization of the bound state of skyrmions
possessing originally characteristic torus-like form [\cite{7}] allows
to obtain the deuteron and the singlet $NN$-scattering state with isospin
$I=1$ [\cite{8}]. The binding energy of the deuteron is about $\sim 30$~MeV
when zero modes quantum corrections are taken into account. It decreases to
$\sim 6$~MeV when some nonzero modes quantum corrections are included
[\cite{9}]. Some states
with positive parity and unusual connection between isospin and angular
momentum are predicted also, which can reveal
themselves as supernarrow radiatively decaying dibaryon [\cite{10}], as well
as dibaryons with negative parity. The experimental indications on the
existence of such states have been obtained recently [\cite{11,12}].
Another issue is the binding energy of strange hypernuclei with atomic number
up to $\sim 16$ which can be described qualitatively under some natural
assumptions, and predictions of more bound charmed or beautiful hypernuclei
[\cite{13}].

Here we shall discuss the nuclides with baryon number $A=B \leq 32$:
mass splittings of isotopes with different $N, Z$, including neutron rich
nuclides with the difference $N-Z$ up to 10 and 11, $N,\;Z$ are the numbers
of neutrons and protons in the nucleus, $N+Z=A$.
Some extrapolation and modification of the approach applied previously
[\cite{4,5,13}] allows to make certain predictions and conclusions also in this case.
In particular, the symmetry energy term in the Weizsacker---Bethe---Bacher mass
formula for nuclei [\cite{14}], $E_{\rm sym} = b_{\rm sym} (N-Z)^2/2A$ can be described
successfully ($b_{\rm sym} \simeq 50$~MeV).
The presence of this term (more correctly, it should be called the asymmetry
energy since it appears due to difference of $N$ and
$Z$ numbers) combined with the energy of Coulomb repulsion
of protons, $E_{\rm C} \simeq 0.6 Z^2e^2/R_Z$, leads to the observed excess of
neutrons in nuclei (see, e.g. [\cite{15,16}]). Within the chiral soliton approach
this symmetry energy originates from isotopical rotations of multiskyrmions
and has the form $E_{\rm sym}= I(I+1)/(2\Theta_I)$, with isospin $I=(N-Z)/2$ for
the states of lowest energy (we shall consider mainly the case $N>Z$, as it
follows from presence of the Coulomb repulsion).
Within the conventional approach this term appears due to dependence of the
Fermi-motion kinetic energies $\epsilon_Z,\; \epsilon_N$ of protons and
neutrons on $N$ and $Z$ numbers [\cite{15}]:
$$ b_{\rm sym} = {2\over 3} (\epsilon_Z+\epsilon_N), $$
$\epsilon_Z \simeq k_{\rm F}(p)^2/2M$, with Fermi-momentum of protons in the nucleus
$k_{\rm F}(p)=(3\pi^2Z/V_p)^{1/3}$, $V_p$~being the volume occupied by protons in the
nucleus, and similarly for $\epsilon_N$ ($V_n$ can be different from $V_p$).
Numerically this gives for heavy nuclei $b_{\rm sym} \simeq 24$~MeV if the radius
of the nucleus (for protons and neutrons) $R_A = 1.12 \,A^{1/3}$~fm,
and approximately the same contribution give potential terms [\cite{15}].
The presence of the linear in isospin $I$ term shifts slightly the balance of
energies $E_{\rm sym}$ and $E_{\rm C}$ towards larger values of $N$, in better agreement
with data, especially for smaller $A$.

The question, if the neutron rich light nuclides may exist, which we address
here as well, seems to be more viable now when the beams of nuclides become available,
and the possibilities
for production and observation of such states have been increased
[\cite{17}$-$\cite{20}].
After short theoretical introduction in the next section we discuss
qualitatively situation with the mass splittings of relatively light nuclear
isotopes (Section 3), give the results for differences of binding energies
of isotopes with integer isospin (Section 4) and half-integer isospin, odd
atomic numbers (Section 5). Some predictions for binding energies of neutron
rich nuclides are given in Sections 4 and 5. Discussion of our method and results
is presented in concluding Section 6.
\section{2. Theoretical background}
The brief description of the chiral soliton approach and some basic formulas
are necessary.
The $SU(2)$ skyrmions are described by the Lagrangian density of the effective
field theory, depending on the
$SU(2)$ matrix $U=f_0+i(\tau_1f_1+\tau_2f_2+\tau_3f_3)$. The
Lagrangian in its ``minimal'' form, suggested by Skyrme [\cite{1}], is
$$ L=-\frac{F_{\pi}^2}{16}{\rm Tr}L_\mu L^\mu +{1 \over 32e^2}{\rm Tr} [L_\mu L_\nu]^2 +
\frac{F_{\pi}^2 m_{\pi}^2} {16} {\rm Tr}(U+U^{\dagger}-2) \eqno (1) $$
with $L_\mu = \partial_\mu U U^{\dagger}=i{\bf l}_\mu \cdot \boldtau$.

The sixth order in chiral derivatives term also can be included into
consideration [\cite{21}]\footnote{Probably, one of the first attempts to
include the sixth order term was made in [\cite{7}] where the bound $B=2$
torus-like configuration was found, similar to the case of fourth order, or
Skyrme term.}:
$$L_6 = - c_6 {\rm Tr}\biggl([L_\mu L^\nu][L_\nu L^\gamma] [L_\gamma L^\mu]\biggr) $$
with dimensional constant $c_6$.
The important property of the Lagrangian $(1)$ is that it is proportional to
the number of colors $N_c$ of the underlying quantum chromodynamics --- theory of
colored quarks and gluons and their interactions [\cite{22}]. The effective
field theory described by Lagrangian density $(1)$ is being built from this
QCD.

The baryon, or winding, number is the fourth
component of the Noether current generated by the Wess-Zumino term
in the action [\cite{22}], and in these notations equals to
$$ B= -{1 \over 4\pi^2} \int {\rm Tr} (L_1L_2L_3) d^3r =
-{1 \over 2\pi^2} \int ({\bf l}_1{\bf l}_2{\bf l}_3) d^3r, \eqno (2) $$
$({\bf a} {\bf b} {\bf c})$ denotes the mixed product of vectors
${\bf a},{\bf b}$, and ${\bf c}$.
We shall rewrite it in less conventional but  transparent form
 $$ B= -\frac{1}{2\pi^2} \int (\vec\partial f_1 \vec\partial f_2
 \vec\partial f_3)/f_0 d^3r = $$
$$=- \frac{1}{\pi^2}  \int \delta (f_0^2+f_1^2+f_2^2+f_3^2-1)
(\vec\partial f_1 \vec\partial f_2 \vec\partial f_3)df_0 d^3r.
\eqno(3) $$
Three functions are introduced usually to describe $f_i$: $f_0=c_F$, $f_k=s_Fn_k$,
where the components of the unit vector $\bf n$ are
$n_1=s_{\alpha}c_{\beta}$, $n_2=s_{\alpha}s_{\beta}$,
$n_3=c_{\alpha}$, $c_F=\cos F$, $s_F=\sin F$, $s_\alpha = \sin\alpha$, etc.
Then $(\vec\partial f_1\vec\partial f_2\vec\partial f_3)/f_0=
s_F^2s_{\alpha} (\vec\partial F \vec\partial\alpha
\vec\partial\beta)$, and the winding number
$$ B= -{1\over 2\pi^2} \int s_F^2 s_\alpha I[(F,\alpha,\beta)/(x,y,z)] d^3r
\eqno (4) $$
with $I[(F, \alpha,\beta)/(x,y,z)]$ --- Jacobian of the transformation from
variables ${\bf r} = x,y,z$ to $F,\,\alpha,\,\beta$. Since the element of
the unit 3-dimensional sphere $dS^3=s_F^2s_\alpha dF\,d\alpha\,d\beta$,
and $2\pi^2$ is the surface of the unit sphere, Eq. $(4)$ shows how many
times the sphere $S^3$ (homeomorphic to $SU(2)$) is covered when integration
over $d^3r$ is made, so it is the degree of the map $R^3 \to SU(2)$.

The static energy of arbitrary $SU(2)$ skyrmion in these notations is
$$M= \int \Biggl\{ \frac{F_{\pi}^2}{8}[(\vec\partial F)^2+s_F^2
((\vec\partial\alpha)^2+
s_{\alpha}^2(\vec\partial\beta)^2)]+ {} $$
$$ {} +\frac{s_F^2}{2e^2} ( [\vec\partial F \times \vec\partial\alpha]^2+
s_{\alpha}^2[\vec\partial F \times \vec\partial\beta]^2+s_F^2s_{\alpha}^2
[\vec\partial\alpha \times \vec\partial\beta]^2)+m_{\pi}^2F_{\pi}^2(1-c_F)/4 + {} $$

$$ {} + 96 c_6 s_F^4s_\alpha^2 (\vec\partial F \vec\partial\alpha
\vec\partial\beta )^2 \Biggr\} d^3r, \eqno (5) $$
$[{\bf a}\times{\bf b}]$ means the vector product of ${\bf a}$ and ${\bf b}$.
The mass $(5)$ is proportional to the number of colors of underlying QCD,
i.e. it is large at large $N_c$.

The rational map approximation (RM) proposed in [\cite{23}] is very effective
for description of multibaryons with large enough baryon numbers. The direct
numerical calculation of configurations of minimal energy [\cite{24}] provides
the results very close to those obtained within the RM approximation.
The profile function $F$ within the RM approximation depends on variable $r$
only, and the unit vector $\bf n$ --- on angular variables $\theta, \phi$:
$n_1= 2 {\rm Re}\,R/(1+|R|^2)$, $n_2=2 {\rm Im}\,R/(1+|R|^2)$ and $n_3= (1-|R|^2)/(1+|R|^2)$,
$R(\theta,\,\phi)$ being a rational function of variable
$z=\tan(\theta/2)\exp(i\phi)$ defining the map of degree ${\cal N}$ of the
2-dimensional spheres $S^2 \to S^2$.
We shall use the following notations, introduced in [\cite{23}]:
$$ {\cal N} = {1\over 8\pi} \int r^2 (\partial_i {\bf n})^2 d\Omega =
{1\over 4\pi} \int \frac{2i dRd\bar{R}}{(1+|R|^2)^2} , \eqno(6) $$
$${\cal I} = {1\over 4\pi} \int r^4 \frac{[\vec\partial n_1 \times
\vec\partial n_2]^2}{n_3^2}d\Omega = {1\over 4\pi} \int
\Biggl[\frac{1+|z|^2}{1+|R|^2}\frac{|dR|}{d|z|}\Biggr]^4
\frac{2idzd\bar{z}}{(1+|z|^2)^2}, \eqno (7) $$
$d\Omega = \sin\theta d\theta d\phi = 2idz d\bar{z}/(1+|z|^2)^2$. For
configurations of lowest energy ${\cal N}=B$ and $F(0)-F(\infty)=\pi$.
The classical mass of the configuration for the RM ansatz can be written as
[\cite{23,21}]
$$ M_{\rm RM} = {\pi F_\pi \over e'} \int \biggl\{r'^2 F'^2 + 2B s_F^2[1+
(1-\lambda)F'^2 ] +{\cal I}{s_F^4\over r^2}\biggl
[1+\lambda(F'^2-1)\biggr]\biggr\}dr + {\rm M.t.}, \eqno(8) $$
where the distance $r$ is measured in units $2/(F_\pi e')$,
$e'=e\sqrt{1-\lambda}$, $\lambda$ defines the weight of the sixth order term
in the Lagrangian according to the relation $\lambda/(1-\lambda)^2 =
48 c_6 F_\pi^2 e^4$. If $\lambda=0$, $c_6=0$ and we obtain original variant of
the Skyrme model (SK4 variant), $\lambda =1$ corresponds to the pure SK6 variant.
In this case relation takes place: $e'=1/(48 c_6F_\pi^2)^{1/4}$.
$\rm M.t.$ denotes the contribution of the mass term.

The energy of the system depending
on the angular velocities of rotations in $SU(2)$ isospin collective
coordinate space can be written in such form:
$$ L_{\rm rot} = \biggl(\frac{ {F_\pi}^2 }{16}+\frac{{\bf l}_1^2+{\bf l}_2^2
+{\bf l}_3^2}{8e^2}\biggr)
 (\tilde{\omega}_1^2+\tilde{\omega}_2^2 + \tilde{\omega}_3^2) +
\frac{(\tilde{\omega}_1{\bf l}_1+\tilde{\omega}_2{\bf l}_2+\tilde{\omega}_3
{\bf l}_3)^2}{8e^2}. \eqno (9) $$
The angular velocities of rotation in the isospace are defined in standard way:
$A^{\dagger}\dot{A}= -{i \over 2}\boldomega\cdot\boldtau$.
The functions $\tilde{\omega}_i$ are connected with the body fixed
angular velocities of $SU(2)$ rotations by means of transformation:
$$ \tilde{\boldomega}\cdot\boldtau=
U^{\dagger}\boldomega\cdot\boldtau U - \boldomega\cdot\boldtau, \eqno (10) $$
with ${\bf l}_1^2+{\bf l}_2^2+{\bf l}_3^2=
(\vec\partial f_0)^2+(\vec\partial f_1)^2+(\vec\partial f_2)^2+
(\vec\partial f_4)^2$ and $\tilde{\boldomega}^2=4[\boldomega^2{\bf f}^2-
(\boldomega\cdot{\bf f})^2]$.
The coefficients in the quadratic form $(9)$ in $\omega_i\omega_k$ define the
$SU(2)$ (isotopic) moments of inertia of arbitrary systems of $SU(2)$ skyrmions.
In the axially symmetrical case and in RM approximation they are not
complicated and were written explicitly in [\cite{8,25,26}].

Some further complication is necessary because
one should introduce another set of collective coordinates [\cite{8}]
to take into account the usual space rotations which we shall describe by
angular velocities $\Omega_i$, $i=1,2,3$.
The part of the Lagrangian describing both kinds of rotations can be written
then [\cite{8}]:
$$ L_{\rm rot}={1 \over 2}U_{ij}\omega_i\omega_j+{1 \over 2}V_{ij}\Omega_i\Omega_j
-W_{ij} \omega_i \Omega_j, \eqno (11) $$
where tensors $U_{ij}$, $V_{ij}$, and $W_{ij}$ are defined by the configuration
of lowest energy for each value of $B$, i.e. they are functions of $F,\alpha$,
and $\beta$.

The body fixed isospin and spin angular momentum operators
$K_l=\partial L_{\rm rot}/\partial \omega_l$ and $L_l=\partial L_{\rm rot}/\partial
\Omega_l$ are linear combinations
of $\omega_i$ and $\Omega_i$.
Three diagonal moments of inertia and three off-diagonal define each of three
rotation terms in $(11)$ in most general case.
In the case of axially symmetric systems we obtained four different
diagonal moments of inertia: $\Theta_1=\Theta_2=\Theta_N$; $\Theta_3$;
$\Theta_R$ and $\Theta_{\rm int}$ [\cite{8}]\footnote{For the hedgehog-type
configurations the isospin and usual space tensors of
inertia coincide since iso- and usual space rotations are identical for the
hedgehogs, and the configuration is described by one common moment of inertia.}.
The static masses of solitons and momenta of inertia were calculated previously
for $B \leq 32$ starting from rational map ansatz [\cite{23,24}] and using
the variational minimization program [\cite{25,26}], see Table 1.
It should be noted that expressions $(1)$$-$$(9)$ contain all
information which is necessary for the description of arbitrary $SU(2)$
skyrmions, in such minimal form. It was sufficient for the fitting
of the mass differences of the nucleon and $\Delta$ isobar [\cite{2,3}] and for
the description of many basic properties of light nuclei [\cite{4,5}].

The moments of inertia, isotopical and orbital, necessary for our purpose here,
are given by the expressions:
$$\Theta_I = {1\over 3}\Theta_{I,aa}= $$
$$= {2\pi\over 3}\int s_F^2\Biggl\{ F_\pi^2 +{4\over e'^2}\Biggl[(1-\lambda)
\Biggl(f'^2 + {\cal N}{s_F^2\over r^2}\Biggr) +4\lambda {\cal N} f'^2
{s_f^2\over (F_\pi e'r)^2} \Biggr]\Biggr\}r^2 dr; \eqno (12)$$

$$\Theta_J= {1\over 3}\Theta_{J,aa}= $$
$$= {2\pi\over 3}\int s_F^2\Biggl\{ {\cal N} F_\pi^2 +{4\over e'^2}\Biggl[(1-\lambda)
\Biggl( {\cal N}f'^2 + {\cal I}{s_F^2\over r^2}\Biggr) +4\lambda {\cal I} f'^2
{s_f^2\over (F_\pi e'r)^2} \Biggr]\Biggr\}r^2 dr. \eqno (13)$$
The inequalities take place for any value of $\lambda$ [\cite{25,26}]:
$$ {{\cal I}\over B} \Theta_I \geq \Theta_J \geq B\Theta_I, \eqno (14) $$
which ensures that the quantum correction due to usual space collective
rotations is always smaller (even much smaller at large $A=B$) than isotopical
quantum correction.
The moments of inertia $(12)$, $(13)$ are proportional
to the number of colors $N_c$,
similar to the classical mass of solitons, therefore, the rotational quantum
corrections are proportional to $\sim 1/N_c$, i.e. they are small, and the
whole consideration becomes selfconsistent at large $N_c$.
The fact that $N_c=3$ in our world makes the real life somewhat more
complicated.

Considerable simplification takes place because of the symmetry properties of
many multiskyrmion configurations, beginning with  $B=3,\;7$ and some other
configurations found in [\cite{23,24}]. It turned out that all three tensors of
inertia in $(11)$ at large $B$ approximately are proportional to the unit
matrix, i.e.
$U_{ij}\simeq u \delta_{ij}$, $V_{ij}\simeq v \delta_{ij}$, $W_{ij}\simeq
w \delta_{ij}$.
Moreover, in many cases of interest the interference tensor of inertia is
negligibly small, $w \ll u$ and $w \ll v$ [\cite{26,27}]; the relative magnitude
of this interference inertia decreases with increasing baryon number, and we
shall neglect it for numerical estimates.

The mass splitting between nucleon and $\Delta(1232)$ --- usually one of the first
quantities fitted within the chiral soliton approach --- is described almost
equally for different choices of the parameters used in the literature
[\cite{2,3,26}] (it scales like $\sim F_\pi e^3$ approximately), but
dimensions of solitons are different since they scale like $1 / (F_{\pi} e)$.
One of the possible choices of the parameters of the model is [\cite{28}],
$F_{\pi}=186$~MeV, $e=4.12$, and we shall use these numbers for the ``nucleon''
variant of the model.
Dimensions of multiskyrmions, the classical field configurations, are much
smaller than dimensions of observable quantized states, i.e. nuclei. The latter
are reproduced only
when the vibration and breathing motions in the solitons are taken into
account, as it was shown recently for the deuteron in [\cite{9}].
The numerical values of inertia are given in Table 1 for the SK4
$(e=4.12 $ and $e=3.0$, marked with $^*$) and SK6 variants of the model, $e'=4.11$
and $2.84$, also marked with $^*$. The values $e=3.0$ and $e'=2.84$ for the
SK6 variant allow to describe well the isotopical splittings of nuclei with
atomic numbers between $\sim 10$ and $ 32$, see discussion in the
following sections.
\begin{center}
\begin{tabular}{|l|l|l|l|l|l|l|}
\hline
$B$& $\Theta_I({\rm SK4})$&$\Theta_J({\rm SK4})$&$\Theta_I({\rm SK6})$&$\Theta_J({\rm SK6})$&
$\Theta_I({\rm SK4})^*$&$\Theta_I({\rm SK6})^*$\\
\hline
$1 $ & $5.55$ & $5.55$ & $5.13$ & $5.13$ & $12.8$ & $14.2$ \\ \hline
$2 $ & $10.4$ & $22.9$ & $9.26$ & $21.9$ & $24.3$ & $25.7$ \\ \hline
$3 $ & $14.8$ & $49.7$ & $12.7$ & $46.0$ & $34.7$ & $35.5$ \\ \hline
$4 $ & $18.2$ & $78.3$ & $15.2$ & $68.8$ & $42.9$ & $43.2$ \\ \hline
$5 $ & $22.7$ & $127 $ & $18.7$ & $111 $ & $53.5$ & $52.9$ \\ \hline
$6 $ & $26.6$ & $178 $ & $21.7$ & $153 $ & $62.6$ & $61.4$ \\ \hline
$7 $ & $29.5$ & $221 $ & $23.9$ & $186 $ & $69.7$ & $68.1$ \\ \hline
$8 $ & $33.9$ & $298 $ & $27.2$ & $251 $ & $79.9$ & $77.4$ \\ \hline
$9 $ & $37.8$ & $376 $ & $30.2$ & $315 $ & $89.0$ & $85.7$ \\ \hline
$10$ & $41.4$ & $455 $ & $32.9$ & $379 $ & $97.4$ & $93.5$ \\ \hline
$11$ & $45.1$ & $547 $ & $35.8$ & $455 $ & $106 $ & $102 $ \\ \hline
$12$ & $48.5$ & $637 $ & $38.4$ & $526 $ & $114 $ & $109 $ \\ \hline
$13$ & $52.1$ & $737 $ & $41.1$ & $606 $ & $122 $ & $117 $ \\ \hline
$14$ & $56.1$ & $865 $ & $44.3$ & $712 $ & $132 $ & $125 $ \\ \hline
$15$ & $59.8$ & $987 $ & $47.0$ & $811 $ & $140 $ & $133 $ \\ \hline
$16$ & $63.2$ & $1110$ & $49.7$ & $908 $ & $148 $ & $141 $ \\ \hline
$17$ & $66.2$ & $1220$ & $52.1$ & $996 $ & $155 $ & $148 $ \\ \hline
$18$ & $70.3$ & $1380$ & $55.3$ & $1130$ & $164 $ & $156 $ \\ \hline
$19$ & $73.8$ & $1540$ & $58.0$ & $1260$ & $173 $ & $164 $ \\ \hline
$20$ & $77.4$ & $1700$ & $60.9$ & $1390$ & $181 $ & $172 $ \\ \hline
$21$ & $80.8$ & $1860$ & $63.5$ & $1520$ & $189 $ & $179 $ \\ \hline
$22$ & $84.3$ & $2030$ & $66.2$ & $1660$ & $197 $ & $186 $ \\ \hline
$23$ & $88.0$ & $2220$ & $69.0$ & $1810$ & $205 $ & $194 $ \\ \hline
$24$ & $91.3$ & $2400$ & $71.5$ & $1950$ & $213 $ & $202 $ \\ \hline
$25$ & $94.7$ & $2590$ & $74.3$ & $2110$ & $221 $ & $209 $ \\ \hline
$26$ & $98.2$ & $2800$ & $77.0$ & $2280$ & $229 $ & $217 $ \\ \hline
$27$ & $102 $ & $3000$ & $79.7$ & $2440$ & $237 $ & $224 $ \\ \hline
$28$ & $105 $ & $3230$ & $82.5$ & $2630$ & $245 $ & $232 $ \\ \hline
$29$ & $108 $ & $3430$ & $85.0$ & $2790$ & $252 $ & $239 $ \\ \hline
$30$ & $112 $ & $3680$ & $87.9$ & $3000$ & $260 $ & $246 $ \\ \hline
$31$ & $115 $ & $3900$ & $90.4$ & $3180$ & $268 $ & $254 $ \\ \hline
$32$ & $118 $ & $4140$ & $93.0$ & $3360$ & $275 $ & $261 $ \\ \hline
\end{tabular}
\end{center}
{\bf Table 1.} The moments of inertia of multiskyrmions in the SK4 variant of
the model $e=4.12$ and $3$ ($\Theta_I({\rm SK4})^*$), and for
the SK6 variant of the model, $e'=4.11$ and $2.84$ ($\Theta_I({\rm SK6})^*$),
in ${\rm GeV}^{-1}$ \\

\section{3. Multiplets of nuclear isotopes}
To get an idea how the chiral soliton approach can describe the existing
data on the mass splittings of the known nuclear isotopes, we consider the mass
differences of some known nuclei with atomic number beginning with $A=8$.

Consider first the mass difference between isosinglet $^8 {\rm Be}$, $J^P=0^+$,
binding energy (b.e.) $\epsilon =56.5$~MeV, and the components of the
isotriplet $\rm^8Li$$-$$\rm^8B$,
$J^P=2^+$, b.e. $41.28$~MeV and $37.74$~MeV (the experimental
data on total binding energies of nuclei are taken mainly from paper [\cite{29}]).
Using the above formula, we obtain easily
$$ \Delta M(A=8)_{0,1} \;=\; [M(^8{\rm Li})+M(^8{\rm B})]/2-M(^8{\rm Be})\;=\;
{1\over \Theta_{I,8}} +{3\over \Theta_{J,8}}. \eqno(15) $$
In the average of masses of isotopes $^8{\rm Li}$ and $^8{\rm B}$ the number of protons
and neutrons is the same as in $^8{\rm Be}$,  and the average Coulomb energy is also
the same with quite good accuracy.
The theoretical value $\sim 40$~MeV for the SK4 variant of the model should
be compared with $17$~MeV from the data. Only collective motion is taken into
account in $(15)$, and peculiarities of nucleon-nucleon interaction are
totally neglected, e.g. the phenomenologically introduced so called pairing
energy [\cite{15,16}] which increases the binding by $\sim 4$~MeV for $^8{\rm Be}$
and decreases by same value for $^8{\rm Li}$ and $^8{\rm B}$. As a result we should
compare $40$~MeV and $\sim 10$~MeV. So, in this case we can speak only
about qualitative agreement.

For $A=12$ there is isosinglet nucleus $^{12}{\rm C}$ with $J^P=0^+$ and b.e.
$92.16$~MeV and isotriplet components $^{12}{\rm B}$ and $^{12}{\rm N}$ with
$J^P=1^+$, b.e. $79.58$ and $74.04$~MeV,
$$\Delta M(A=12)_{0,1} ={1\over \Theta_{I,12} } +{1\over \Theta_{J,12}}, \eqno(16)$$
which is $22.2$~MeV to be compared with $15.35$~MeV experimentally.
The pairing energy in this case also makes the disagreement even worth by
$5$$-$$6$~MeV.

For $A=16$ isosinglet and scalar nucleus $^{16}{\rm O}$, b.e. $127.62$~MeV
should be compared with nucleus $^{16}{\rm N}$, $J^P=2^-$, b.e. $117.98$~MeV and
$^{16}{\rm F}$, b.e. $111.41$~MeV, if we assume that the last one also has $J^P=2^-$.
In this case
$$\Delta M(A=16)_{0,1} ={1\over \Theta_{I,16}} +{3\over \Theta_{J,16}}, \eqno(17)$$
numerically it is $18.55$~MeV to be compared with $14.69$~MeV, or with
$\sim 10$~MeV when pairing energy is added and subtracted in proper way.

For the case of $A=10$ we have isosinglet $^{10}{\rm B}$, $J^P=3^+$, $\epsilon =
64.75$~MeV and components of isotriplet $^{10}{\rm Be}$ and $^{10}{\rm C}$, $J^P=0^+$,
b.e. $64.98$ and $60.32$~MeV.
Theoretical difference of masses is
$$\Delta M(A=10)_{0,1} ={1\over \Theta_{I,10}} - {6\over \Theta_{J,10}}, \eqno (18)$$
which is about $ 11$~MeV, to be compared with experimental value $ 2.1$~MeV.
The pairing energy of the order of $3$$-$$4$~MeV decreases the binding of
$^{10}{\rm B} \;(N=Z=5)$ and increases the binding of $^{10}{\rm Be}$
and $^{10}{\rm C}$, so we should compare $11$~MeV and $8$$-$$10$~MeV.

To summarize, we can state that the agreement between data and theory is only
qualitative for small $A=B$, if we take the value of the model parameter
$e=4.12$ which allowed to describe the mass difference between nucleons and
$\Delta (1232)$ isobar (we shall call this as ``nucleon'' variant of the model),
which improves however with increasing atomic number, as can be seen for
$A=12, 16$.

It is clear that the ``nucleon'' variant of the model cannot be good in the case
of nuclei: the chiral field configurations given by rational map ansatz,
although confirmed by direct numerical computation, represent the ``sceleton''
configurations which differ from realistic distribution of nucleons inside of
nuclei. The nonzero mode motion should be taken into account --- vibration,
breathing, etc. to get more realistic distribution of matter inside of
multibaryons. Technically, this problem is very complicated. Up to now, the
mathematical analysis of these modes was made for baryonic numbers up to
$A=7$ [\cite{30}],
and some numerical estimates were made for the deuteron $(A=2)$ [\cite{9}], as it
was mentioned in Introduction. For each value of $A$ we can improve the
agreement of theoretical estimates with data if we allow the only parameter
of the model, Skyrme constant $e$, to vary --- to decrease, in fact, in order to
make the dimensions of multiskyrmion and moments of inertia greater.
Effectively, such approach allows to take into account the vibration and
breathing of separate parts of multiskyrmions. Evidently, for each value of
atomic number we can get perfect agreement if $e=e(A)$. It is, however, more
instructive to describe a number of nuclei with the same value of constant $e$.
The optimal value of the constant is $e=3.0$ (SK4 variant) and $e=2.84$, as
analysis of data shows (next section).

With rescaled value of the constant $e$ we obtain (see Table 1):
$\Delta M(A=8) = 16.8$~MeV; $\Delta M(A=12) = 9.5$~MeV and $\Delta M(A=16) =
7.9$~MeV, and for $A=12$ and $16$ there is now agreement with data within
$2$$-$$3$~MeV. As we shall see below, quite satisfactory agreement with data can
be obtained when appropriate choice of nuclear isotopes is made, for $A$
between $\sim 10$ and $\sim 30$.
\section{4. Even baryon numbers, integer isospins}

It is possible to obtain more adequate comparison of the model predictions with
data, comparing the differences of masses of nuclei with even isospins, odd
isospins, and similar differences for half-integer isospins (next section).
As it was mentioned in previous section, the  nucleons pairing energy should
be added to the binding energy, which equals to $\Delta$ when both $N,\;Z$ are
even, $0$ when $A$ is odd, and $-\Delta$ when both $N,\;Z$ are odd
(see, e.g. [\cite{15}]). This pairing energy decreases with increasing atomic
number; it means that peculiarities of nucleon interaction become less important
for large $A$.
In the differences of masses of isotopes with the same pairing energy this
specific contribution is cancelled, and one can hope that the collective
motion effects --- which are mainly taken into account within the chiral soliton
approach --- play the dominant role.

Formulas for energies differences can be correctly applied
to the same nucleus' quantum states with different isospin.
However, experimental data for excited states with higher isospin
are lacking often. Nevertheless, using the isotopical invariance we can obtain
binding energies of this level subtracting the Coulomb energy difference
from binding energies of the components of isotopical multiplet with neutron
excess.

In Tables 2$-$10 the Coulomb energies are calculated
according to [\cite{16}] (finite-range liquid-drop model):
$$ E_{\rm C} = \frac{3}{5} \frac{e^2 Z^2}{r_0 A^{1/3}} B_3 -
   \frac{3}{4} \left(\frac{3}{2\pi}\right)^{2/3} \frac{e^2 Z^{4/3}}{r_0 A^{1/3}}
-
   \frac{1}{8} \left(\frac{145}{48}\right) \frac{r_p^2 e^2 Z^2}{r_0^3 A},
\eqno (19) $$
where nuclear-radius constant $r_0 = 1.16$~fm,
proton root-mean-square radius $r_p = 0.80$~fm.
The first term in $(19)$ is Coulomb energy for a homogeneously charged,
diffuse-surface nucleus to all orders in the diffuseness
constant $a_{\rm den} ,$ the second is Coulomb exchange correction,
and the third is proton formfactor correction to the Coulomb energy.
The constant
$B_3 = B_3\left(r_0 A^{1/3}/a_{\rm den}\right),$ is normalized so that for zero
diffuseness
$a_{\rm den}$  (range of Yukawa function $\exp(-r/a_{\rm den})$) $B_3=1$. For
spherically symmetrical
nuclei it can be calculated analytically and presented as a sum
$B_3 = 1 - 5 y_0^2+...$, here $y_0=a_{\rm den}/(r_0A^{1/3})$, more details can be
found in [\cite{16}].

The result provided by  $(19)$ for intermediate values of $A$ (practically,
between $A\sim 25$ and $A\sim 50$) can be described by simple formula
$$ E_{\rm C} = \frac{3}{5} \frac{Q^2}{r_{\rm C} A^{1/3}} , \eqno (20) $$
where $Q$ is nucleus's charge, $r_{\rm C} = 1.48$~fm [\cite{14}].
The effective Coulomb radius in $(20)$ is greater than $r_{\rm C}$ given, e.g. in
[\cite{15}],  $r_{\rm C} = 1.24$~fm. The formula $(20)$ is valid
for uniformly charged sphere of radius $R=r_{\rm C} A^{1/3}$ and total charge $Q$.
For other types of charge distribution inside of nuclei the above formula
for $E_{\rm C}$ is modified slightly. For example, for thin shell-like distribution
suggested by rational map approximation for multiskyrmions, the Coulomb
energy equals to
$$E_{\rm C} = {1 \over 2} \frac{Q^2}{r_{\rm C} A^{1/3}}, \eqno(21) $$
which is smaller than in the case of uniformly charged sphere by $ 17\%$ only.

For the nuclei considered in Tables 2$-$5 and 7$-$9,
the Coulomb energy increases from zero for nucleus $^6{\rm H}$
to $43.9$~MeV for nucleus $^{32}{\rm S}$.

\begin{center}
\begin{tabular}{|l|l|l|l|l|l|l|}
\hline
$A$&$\epsilon_0^{\rm exp}$&$\epsilon_2^{\rm exp}$&$\Delta\epsilon_{\rm C}$&$\Delta\epsilon_{02}^{\rm exp}$
&$\Delta\epsilon_{02}^{\rm th}$&$\Delta\epsilon_{02}^{\rm th*}$\\ \hline

$\rm^6Li   $$-$$\rm^6H    $ & $32.0 $ & $5.8  $ & $1.1 $ & $27.3$ & $112.9$ & $47.9$ \\ \hline
$\rm^8Be   $$-$$\rm^8He   $ & $56.5 $ & $31.4 $ & $2.0 $ & $27.1$ & $88.5 $ & $37.5$ \\ \hline
$\rm^{10}B $$-$$\rm^{10}Li$ & $64.8 $ & $45.3 $ & $2.9 $ & $22.4$ & $72.5 $ & $30.8$ \\ \hline
$\rm^{12}C $$-$$\rm^{12}Be$ & $92.2 $ & $68.7 $ & $3.8 $ & $27.3$ & $61.8 $ & $26.3$ \\ \hline
$\rm^{14}N $$-$$\rm^{14}B $ & $104.7$ & $85.4 $ & $4.6 $ & $23.9$ & $53.5 $ & $22.8$ \\ \hline
$\rm^{16}O $$-$$\rm^{16}C $ & $127.6$ & $110.8$ & $5.4 $ & $22.2$ & $47.5 $ & $20.3$ \\ \hline
$\rm^{18}F $$-$$\rm^{18}N $ & $137.4$ & $126.5$ & $6.1 $ & $17.0$ & $42.7 $ & $18.2$ \\ \hline
$\rm^{20}Ne$$-$$\rm^{20}O $ & $160.6$ & $151.4$ & $6.8 $ & $16.0$ & $38.7 $ & $16.6$ \\ \hline
$\rm^{22}Na$$-$$\rm^{22}F $ & $174.1$ & $167.7$ & $7.5 $ & $13.9$ & $35.6 $ & $15.2$ \\ \hline
$\rm^{24}Mg$$-$$\rm^{24}Ne$ & $198.3$ & $191.8$ & $8.2 $ & $14.7$ & $32.9 $ & $14.1$ \\ \hline
$\rm^{26}Al$$-$$\rm^{26}Na$ & $211.9$ & $208.2$ & $8.9 $ & $12.6$ & $30.5 $ & $13.1$ \\ \hline
$\rm^{28}Si$$-$$\rm^{28}Mg$ & $236.5$ & $231.6$ & $9.5 $ & $14.4$ & $28.5 $ & $12.3$ \\ \hline
$\rm^{30}P $$-$$\rm^{30}Al$ & $250.6$ & $247.8$ & $10.2$ & $13.0$ & $26.8 $ & $11.5$ \\ \hline
$\rm^{32}S $$-$$\rm^{32}Si$ & $271.8$ & $271.4$ & $10.8$ & $11.2$ & $25.3 $ & $10.9$ \\ \hline
\end{tabular}
\end{center}
{\bf Table 2.} The binding energies differences for isotopes with isospin
$I=0$ and $2$ for the original variant, $e=4.12$, and for the variant with
rescaled constant, $e=3$ (numbers with the $^*$) \\

The pairing energy has different sign for nuclei with $A=6,\;10,\;14,...$
and $A=8,\;12,\;16,...$, but anyway, it is cancelled in the binding energies
differences we consider here.

To obtain the binding energy of isotopes with isospin $I=0$ we take the known
binding energy of ground states of nuclei with $N=Z$ and add the energy of the
Coulomb repulsion given by $(19)$, $(20)$. In the case of $I=2$ we take the ground
state of nucleus with $N=Z+4$ which is supposed to be the component of $I=2$
isomultiplet with the isospin projection $I_3=-2$, and add the Coulomb repulsion
energy as well. The quantity $\Delta \epsilon^{\rm exp}$ shown in all Tables
includes the Coulomb correction.

The energy of usual space rotation of nucleus as a whole, equal to
$J(J+1)/2\Theta_J$, is small for large enough nuclei since $\Theta_J$ is large
(see Table 1), and is omitted within this approach. Another reason to omit
this correction is that the value of $J$ is not known in some cases.

It follows from Table 2 that for light nuclei, $A$ between $6$ and $10$,
there is only qualitative agreement between data and theoretical results, even
with rescaled constant $e$. For atomic numbers between $12$ and $32$ the
agreement is satisfactory and even good in some cases (see column with
$\Delta \epsilon ^{\rm th,*}$. By this reason we shall not consider further small
values of $A$.
\begin{center}
\begin{tabular}{|l|l|l|l|l|l|l|}
\hline
$A$&$\epsilon_1^{\rm exp}$&$\epsilon_3^{\rm exp}$&$\Delta\epsilon_{\rm C}$&
$\Delta\epsilon_{13}^{\rm exp}$
&$\Delta\epsilon_{13}^{\rm th}$&$\Delta\epsilon_{13}^{\rm th*}$\\ \hline

$\rm^{10}Be$$-$$\rm^{10}He$ & $65.0 $ & $30.3 $ & $2.0 $ & $36.7$ & $120.8$ & $51.3$ \\ \hline
$\rm^{12}B $$-$$\rm^{12}Li$ & $79.6 $ & $44.4 $ & $2.9 $ & $38.1$ & $103.0$ & $43.8$ \\ \hline
$\rm^{14}C $$-$$\rm^{14}Be$ & $105.3$ & $70.0 $ & $3.7 $ & $39.0$ & $89.1 $ & $38.0$ \\ \hline
$\rm^{16}N $$-$$\rm^{16}B $ & $118.0$ & $88.2 $ & $4.5 $ & $34.3$ & $79.1 $ & $33.8$ \\ \hline
$\rm^{18}O $$-$$\rm^{18}C $ & $139.8$ & $115.7$ & $5.2 $ & $29.3$ & $71.2 $ & $30.4$ \\ \hline
$\rm^{20}F $$-$$\rm^{20}N $ & $154.4$ & $134.2$ & $6.0 $ & $26.2$ & $64.5 $ & $27.6$ \\ \hline
$\rm^{22}Ne$$-$$\rm^{22}O $ & $177.8$ & $162.0$ & $6.7 $ & $22.5$ & $59.3 $ & $25.4$ \\ \hline
$\rm^{24}Na$$-$$\rm^{24}F $ & $193.5$ & $179.1$ & $7.4 $ & $21.8$ & $54.8 $ & $23.5$ \\ \hline
$\rm^{26}Mg$$-$$\rm^{26}Ne$ & $216.7$ & $201.6$ & $8.1 $ & $23.2$ & $50.9 $ & $21.9$ \\ \hline
$\rm^{28}Al$$-$$\rm^{28}Na$ & $232.7$ & $218.4$ & $8.7 $ & $23.0$ & $47.5 $ & $20.4$ \\ \hline
$\rm^{30}Si$$-$$\rm^{30}Mg$ & $255.6$ & $241.6$ & $9.4 $ & $23.4$ & $44.6 $ & $19.2$ \\ \hline
$\rm^{32}P $$-$$\rm^{32}Al$ & $270.9$ & $259.2$ & $10.0$ & $21.7$ & $42.2 $ & $18.2$ \\ \hline
\end{tabular}
\end{center}
{\bf Table 3.} The binding energies differences for isotopes with isospin
$I=1$ and $3$ for the original variant, $e=4.12$, and for the variant with
rescaled constant, $e=3$ \\

In the case presented in Table 3 we take the binding energies of isotopes
with $N=Z+2$ and $N=Z+6$
(ground states) which are considered as components of isomultiplets $I=1,\;
I_3=-1$ and $I=3, \;I_3=-3$, Coulomb energy is included into consideration
as well. The agreement of theoretical estimates with data is quite satisfactory
for $A$ between $14$ and $32$, qualitative agreement takes place for
$A=10,\;12$ as well.

\begin{center}
\begin{tabular}{|l|l|l|l|l|l|l|}
\hline
$A$&$\epsilon_0^{\rm exp}$&$\epsilon_4^{\rm exp}$&$\Delta\epsilon_{\rm C}$&$\Delta\epsilon_{04}^{\rm exp}$
&$\Delta\epsilon_{04}^{\rm th}$&$\Delta\epsilon_{04}^{\rm th*}$\\ \hline
$\rm^{16}O $$-$$\rm^{16}Be$ & $127.6$ & \ ---   & $9.0 $ & \ ---  & $158.3$ & $67.5$ \\ \hline
$\rm^{18}F $$-$$\rm^{18}B $ & $137.4$ & $\;89.0^\dagger$ & $10.5$ & $59.4$ & $142.3$ & $60.8$ \\ \hline
$\rm^{20}Ne$$-$$\rm^{20}C $ & $160.6$ & $119.2$ & $12.0$ & $53.4$ & $129.1$ & $55.2$ \\ \hline
$\rm^{22}Na$$-$$\rm^{22}N $ & $174.1$ & $140.0$ & $13.4$ & $47.5$ & $118.6$ & $50.8$ \\ \hline
$\rm^{24}Mg$$-$$\rm^{24}O $ & $198.3$ & $168.5$ & $14.8$ & $44.6$ & $109.5$ & $47.0$ \\ \hline
$\rm^{26}Al$$-$$\rm^{26}F $ & $211.9$ & $184.5$ & $16.2$ & $43.6$ & $101.8$ & $43.7$ \\ \hline
$\rm^{28}Si$$-$$\rm^{28}Ne$ & $236.5$ & $206.9$ & $17.5$ & $47.1$ & $95.1 $ & $40.9$ \\ \hline
$\rm^{30}P $$-$$\rm^{30}Na$ & $250.6$ & $224.9$ & $18.8$ & $44.5$ & $89.3 $ & $38.4$ \\ \hline
$\rm^{32}S $$-$$\rm^{32}Mg$ & $271.8$ & $249.7$ & $20.1$ & $42.2$ & $84.4 $ & $36.3$ \\ \hline
\end{tabular}
\end{center}
{\bf Table 4.} The binding energies differences for isotopes with isospin
$I=0$ and $4$ for the original variant, $e=4.12$, and for the variant with
rescaled constant, $e=3$ \\

For $I=4$ the ground states of isotopes with $N=Z+8$ are taken as $I_3=-4$
components of corresponding isomultiplets. The agreement of data and
theoretical values for atomic numbers between $18$ and $30$ is good or
satisfactory, according to Table 4. For the nucleus $^{16}{\rm Be}$ we obtain the
b.e. about $60$~MeV, so it is unstable: it can decay into isotope
$^{14}{\rm Be}$ (b.e. about $70$~MeV) and 2 neutrons.
This result can be considered as prediction of our approach. For the $^{18}{\rm B}$
isotope our prediction for the b.e. is $87.1$~MeV\footnote{The value
$87.1\;\mbox{MeV}=(137.4 - 60.8 + 10.5)\;\mbox{MeV}$, where $60.8$~MeV is the difference of
quantum corrections to the multiskyrmion, rescaled variant, and $10.5$~MeV
is difference of Coulomb energies}, in good agreement with
extrapolation result $89$~MeV of [\cite{29}], marked in Table 4 with
$^\dagger$.

\begin{figure}[h]
\begin{center}
\epsfig{figure=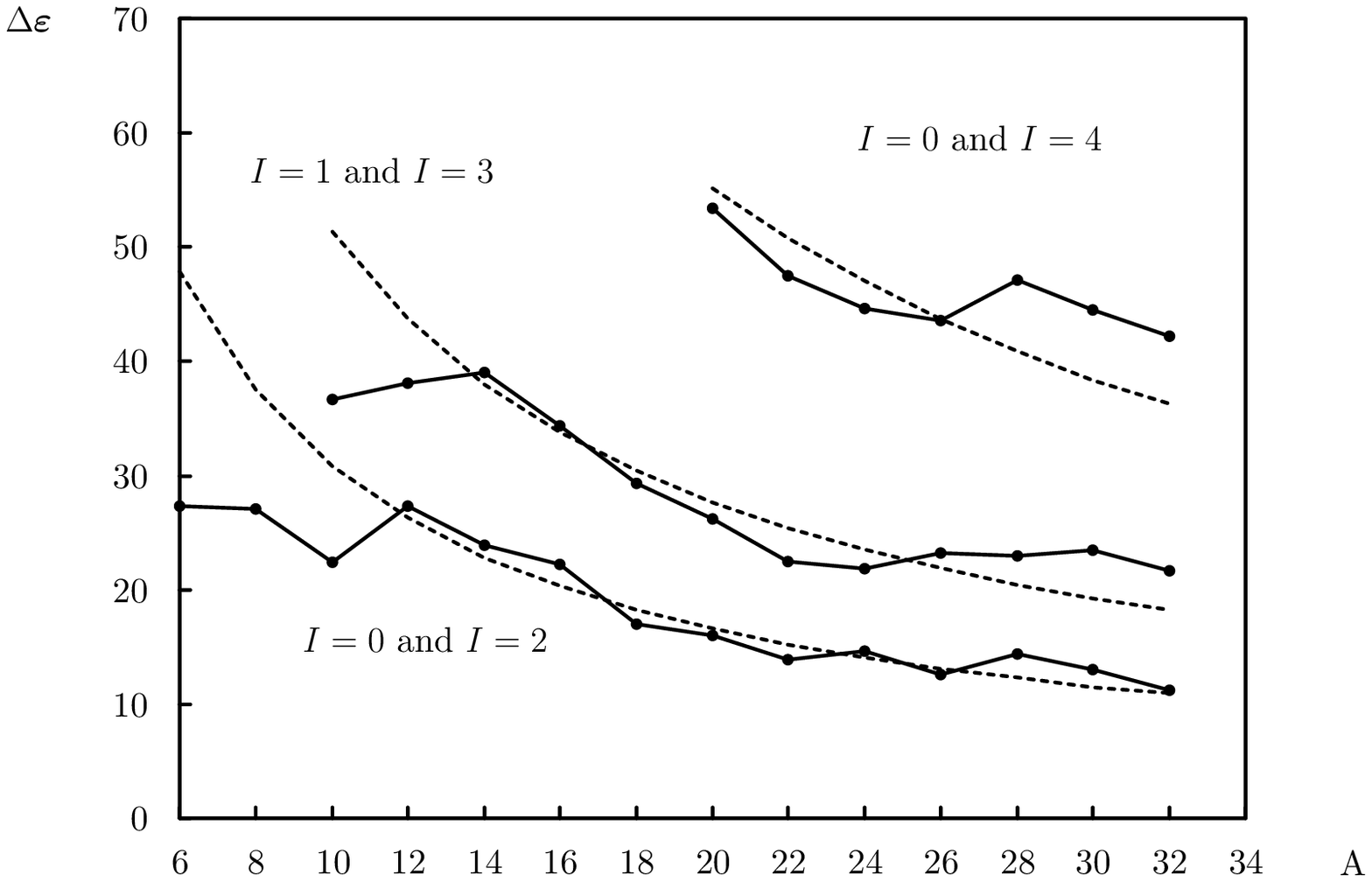,width=16cm,angle=0}
\end{center}
{\bf Fig.\ 1.} The binding energies differences (in MeV)
for isotopes with even atomic numbers for the SK4 variant
with rescaled constant, $e=3.0$
(black points connected with solid lines --- experimental data,
dashed lines --- model calculations)
\end{figure}

Results presented in Tables 2$-$4 are illustrated in Fig.\ 1.
\begin{center}
\begin{tabular}{|l|l|l|l|l|l|l|}
\hline
$A$&$\epsilon_0^{\rm exp}$&$\epsilon_2^{\rm exp}$&$\Delta\epsilon_{\rm C}$&$\Delta\epsilon_{02}^{\rm exp}$
&$\Delta\epsilon_{02}^{\rm SK6}$&$\Delta\epsilon_{02}^{\rm SK6,*}$\\ \hline

$\rm^8Be   $$-$$\rm^8He   $ & $56.5 $ & $31.4 $ & $2.0 $ & $27.1$ & $110.1$ & $38.8$ \\ \hline
$\rm^{10}B $$-$$\rm^{10}Li$ & $64.8 $ & $45.3 $ & $2.9 $ & $22.4$ & $91.0 $ & $32.1$ \\ \hline
$\rm^{12}C $$-$$\rm^{12}Be$ & $92.2 $ & $68.7 $ & $3.8 $ & $27.3$ & $78.0 $ & $27.5$ \\ \hline
$\rm^{14}N $$-$$\rm^{14}B $ & $104.7$ & $85.4 $ & $4.6 $ & $23.9$ & $67.7 $ & $23.9$ \\ \hline
$\rm^{16}O $$-$$\rm^{16}C $ & $127.6$ & $110.8$ & $5.4 $ & $22.2$ & $60.3 $ & $21.3$ \\ \hline
$\rm^{18}F $$-$$\rm^{18}N $ & $137.4$ & $126.5$ & $6.1 $ & $17.0$ & $54.3 $ & $19.2$ \\ \hline
$\rm^{20}Ne$$-$$\rm^{20}O $ & $160.6$ & $151.4$ & $6.8 $ & $16.0$ & $49.3 $ & $17.5$ \\ \hline
$\rm^{22}Na$$-$$\rm^{22}F $ & $174.1$ & $167.7$ & $7.5 $ & $13.9$ & $45.3 $ & $16.1$ \\ \hline
$\rm^{24}Mg$$-$$\rm^{24}Ne$ & $198.3$ & $191.8$ & $8.2 $ & $14.7$ & $41.9 $ & $14.9$ \\ \hline
$\rm^{26}Al$$-$$\rm^{26}Na$ & $211.9$ & $208.2$ & $8.9 $ & $12.6$ & $38.9 $ & $13.8$ \\ \hline
$\rm^{28}Si$$-$$\rm^{28}Mg$ & $236.5$ & $231.6$ & $9.5 $ & $14.4$ & $36.4 $ & $13.0$ \\ \hline
$\rm^{30}P $$-$$\rm^{30}Al$ & $250.6$ & $247.8$ & $10.2$ & $13.0$ & $34.1 $ & $12.2$ \\ \hline
$\rm^{32}S $$-$$\rm^{32}Si$ & $271.8$ & $271.4$ & $10.8$ & $11.2$ & $32.3 $ & $11.5$ \\ \hline
\end{tabular}
\end{center}
{\bf Table 5.} The binding energies differences for isotopes with isospin
$I=0$ and $2$ for the SK6 variant, $e'=4.11,$ and for the SK6 variant with
rescaled constant, $e'=2.84$ \\

In Table 5 we present the comparison of data for differences of binding
energies of isotopes with isospin $0$ and $2$ (as in Table 2) and results
of the SK6 variant of the model, for two values of effective constant $e$.
The rescaling of the constant  $e'$ to the value $2.84$ allows to make agreement
of theoretical prediction and data quite satisfactory, beginning with atomic
number $A=12$, similar to the SK4
variant of the model. We conclude therefore that there is no difference of
principle between both variants of the model, and will make further comparison
with the results provided by the SK4 variant of the model. Our predictions for
the binding energies of nuclides with highest values of isospin (Tables 6, 10)
will be made for both variants of the model.

In view of successful description of binding energies of many isotopes, it
is possible to make predictions of binding energies for neutron rich nuclides
which have greater isospin than we have considered before, e.g.
$I=5$ or $N=Z+10$.
\begin{center}
\begin{tabular}{|l|l|l|l|l|l|l|}
\hline
$A$&$14/\Theta_I^{\rm SK4}$&$\Delta\epsilon_{\rm C}$&$\epsilon_{I=5}^{\rm SK4}$
&$\epsilon_{I=5}^{\rm SK6}$&$\epsilon_{I=5}^\dagger$ &$\epsilon_{I=5}^{\rm exp}$ \\ \hline
$\rm^{18}Be$ & $\;\; 85.4$ & $\;8.8$ & $\;63$& $\;59$ &\ --- & \ ---  \\ \hline
$\rm^{20}B $ & $\;\; 77.3$ & $10.3$ & $\;87$ & $\;83$ &\ --- & \ ---  \\ \hline
$\rm^{22}C $ & $\;\; 71.1$ & $11.8$ & $118$  & $114 $ &\ --- & \ ---  \\ \hline
$\rm^{24}N $ & $\;\; 65.7$ & $13.2$ & $141$  & $138 $ &\ --- & \ ---  \\ \hline
$\rm^{26}O $ & $\;\; 61.1$ & $14.6$ & $170$  & $167 $ &$168$ & \ ---  \\ \hline
$\rm^{28}F $ & $\;\; 57.1$ & $15.9$ & $191$  & $188 $ &$186$ & \ ---  \\ \hline
$\rm^{30}Ne$ & $\;\; 53.8$ & $17.2$ & $219$  & $216 $ &\ --- & $212  $\\ \hline
$\rm^{32}Na$ & $\;\; 50.9$ & $18.5$ & $238$  & $235 $ &\ --- & $231  $\\ \hline
\end{tabular}
\end{center}
{\bf Table 6.} Predictions for the binding energies of neutron rich nuclides
with $N=Z+10$ or $I=5$ for the SK4 and SK6 variants of the model.
The values $\epsilon^\dagger$ are the results of extrapolation of [\cite{29}] \\

The binding energy of neutron rich nuclides with $N=Z+10$ shown in
Table 6 is calculated according to the following formula:
$$ \epsilon_{N=Z+10} = \epsilon_{N=Z+2} - {14 \over \Theta_{A=N+Z}} +
\Delta\epsilon_{\rm C}, \eqno (22) $$
where $\epsilon_{N=Z+2}$ is the experimentally known binding energy of the isotopes with
isospin $I=1$, the number $14$ is half of difference of Casimir operator $I(I+1)$
for $I=5$ and $I=1$,  $\Delta\epsilon_{\rm C}$ is the difference of the
repulsive Coulomb energies for nuclides with the same $A=N+Z$ and different $Z$,
calculated with formula $(19)$. Predictions of two variants of the
model do not differ much, the maximal difference is about $4$~MeV for lighter
nuclides.
\section{5. Odd baryon numbers, half-integer isospin}
The consideration similar to that made for integer isospin in previous section
can be performed for half-integer isospins, i.e. for
nuclei with odd atomic numbers. The phenomenologically introduced pairing
energy is absent in this case.
\begin{center}
\begin{tabular}{|l|l|l|l|l|l|l|}
\hline
$A$&$\epsilon_{1/2}^{\rm exp}$&$\epsilon_{5/2}^{\rm exp}$&$\Delta\epsilon_{\rm C}$&
$\Delta\epsilon_{1/2,5/2}^{\rm exp}$
&$\Delta\epsilon_{1/2,5/2}^{\rm th}$&$\Delta\epsilon_{1/2,5/2}^{\rm th*}$\\ \hline
$\rm^9Be   $$-$$\rm^9He   $ & $58.2 $ & $30.3 $ & $2.0 $ & $29.9$ & $105.8$ & $45.0$ \\ \hline
$\rm^{11}B $$-$$\rm^{11}Li$ & $76.2 $ & $45.6 $ & $2.9 $ & $33.5$ & $88.6 $ & $37.7$ \\ \hline
$\rm^{13}C $$-$$\rm^{13}Be$ & $97.1 $ & $68.1 $ & $3.7 $ & $32.7$ & $76.8 $ & $32.7$ \\ \hline
$\rm^{15}N $$-$$\rm^{15}B $ & $115.5$ & $88.2 $ & $4.5 $ & $31.8$ & $66.9 $ & $28.6$ \\ \hline
$\rm^{17}O $$-$$\rm^{17}C $ & $131.8$ & $111.5$ & $5.3 $ & $25.6$ & $60.4 $ & $25.8$ \\ \hline
$\rm^{19}F $$-$$\rm^{19}N $ & $147.8$ & $132.0$ & $6.0 $ & $21.8$ & $54.1 $ & $23.2$ \\ \hline
$\rm^{21}Ne$$-$$\rm^{21}O $ & $167.4$ & $155.2$ & $6.8 $ & $19.0$ & $49.4 $ & $21.2$ \\ \hline
$\rm^{23}Na$$-$$\rm^{23}F $ & $186.6$ & $175.3$ & $7.5 $ & $18.8$ & $45.5 $ & $19.5$ \\ \hline
$\rm^{25}Mg$$-$$\rm^{25}Ne$ & $205.6$ & $196.0$ & $8.2 $ & $17.8$ & $42.3 $ & $18.1$ \\ \hline
$\rm^{27}Al$$-$$\rm^{27}Na$ & $225.0$ & $214.9$ & $8.8 $ & $18.9$ & $39.4 $ & $16.9$ \\ \hline
$\rm^{29}Si$$-$$\rm^{29}Mg$ & $245.0$ & $235.3$ & $9.5 $ & $19.2$ & $36.9 $ & $15.9$ \\ \hline
$\rm^{31}P $$-$$\rm^{31}Al$ & $262.9$ & $255.0$ & $10.1$ & $18.0$ & $34.7 $ & $14.9$ \\ \hline
\end{tabular}
\end{center}
{\bf Table 7.} The binding energies differences for isotopes with isospin
$I=1/2$ and $5/2$ for the original variant, $e=4.12$, and for the variant with
rescaled constant, $e=3$ \\

In Table 7 we take the binding energies of nuclei with $N=Z+1$ for isospin
$I=1/2$ and $N=Z+5$ for $I=5/2$. For the rescaled variant of the model
the quantitative agreement with data is quite good for atomic numbers between
$13$ and $25$, qualitative agreement takes place for all values of $A$
presented in Table 7.
\begin{center}
\begin{tabular}{|l|l|l|l|l|l|l|}
\hline
$A$&$\epsilon_{3/2}^{\rm exp}$&$\epsilon_{7/2}^{\rm exp}$&$\Delta\epsilon_{\rm C}$&
$\Delta\epsilon_{3/2,7/2}^{\rm exp}$
&$\Delta\epsilon_{3/2,7/2}^{\rm th}$&$\Delta\epsilon_{3/2,7/2}^{\rm th*}$\\ \hline
$\rm^{15}C $$-$$\rm^{15}Be$ & $106.5$ & \ ---   & $3.7 $ & \ ---  & $100.4$ & $42.8$ \\ \hline
$\rm^{17}N $$-$$\rm^{17}B $ & $123.9$ & $89.6 $ & $4.4 $ & $38.7$ & $90.6 $ & $38.7$ \\ \hline
$\rm^{19}O $$-$$\rm^{19}C $ & $143.8$ & $115.8$ & $5.2 $ & $33.2$ & $81.2 $ & $34.7$ \\ \hline
$\rm^{21}F $$-$$\rm^{21}N $ & $162.5$ & $138.8$ & $5.9 $ & $29.6$ & $74.1 $ & $31.8$ \\ \hline
$\rm^{23}Ne$$-$$\rm^{23}O $ & $183.0$ & $164.8$ & $6.6 $ & $24.8$ & $68.2 $ & $29.2$ \\ \hline
$\rm^{25}Na$$-$$\rm^{25}F $ & $202.5$ & $183.5$ & $7.3 $ & $25.3$ & $63.4 $ & $27.2$ \\ \hline
$\rm^{27}Mg$$-$$\rm^{27}Ne$ & $223.1$ & $203.0$ & $8.0 $ & $28.1$ & $59.0 $ & $25.4$ \\ \hline
$\rm^{29}Al$$-$$\rm^{29}Na$ & $242.1$ & $222.8$ & $8.7 $ & $28.0$ & $55.4 $ & $23.8$ \\ \hline
$\rm^{31}Si$$-$$\rm^{31}Mg$ & $262.2$ & $244.0$ & $9.3 $ & $27.5$ & $52.1 $ & $22.4$ \\ \hline
\end{tabular}
\end{center}
{\bf Table 8.} The binding energies differences for isotopes with isospin
$I=3/2$ and $7/2$ for the original variant, $e=4.12$, and for the variant with
rescaled constant, $e=3$ \\

The nuclei with $N=Z+3$ and $N=Z+7$ in Table 8 correspond
to the states with isospin $I=3/2,\;I_3=-3/2$ and $I=7/2,\;I_3=-7/2$.
Good agreement between data and theory (rescaled variant) takes place
between $A=17$ and $27$.
The nucleus $^{15}{\rm Be}$ is not observed yet, so, the b.e. of
this nucleus $63.7$~MeV is prediction of the model. It is not stable since it
can decay into $^{14}{\rm Be}$ (b.e. $70$~MeV) plus neutron.
\begin{figure}
\begin{center}
\epsfig{figure=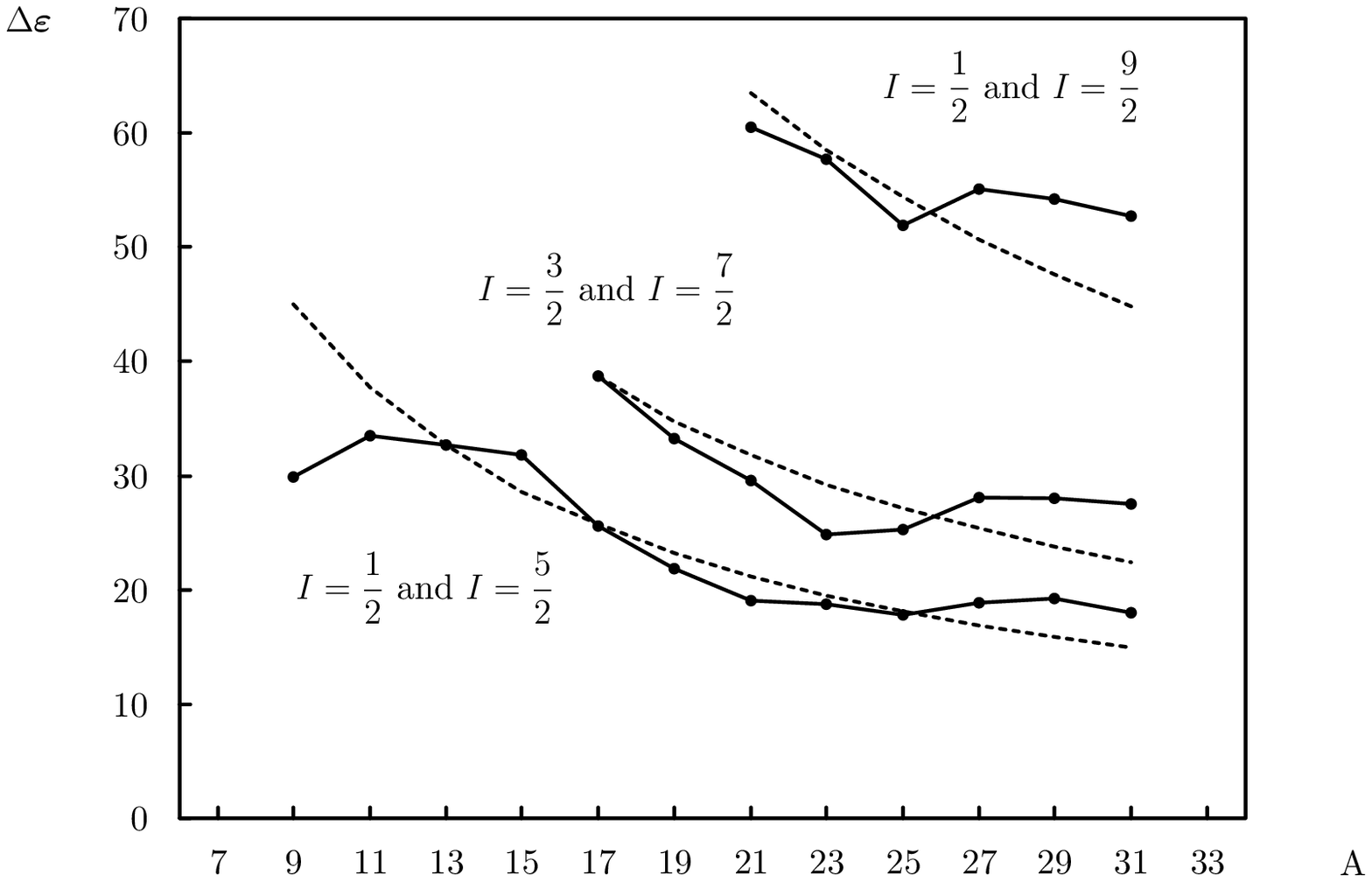,width=16cm,angle=0}
\end{center}
{\bf Fig.\ 2.} The binding energies differences (in MeV) for isotopes
with odd atomic numbers for the rescaled SK4 variant of the model, $e=3.0$
(black points connected with solid lines --- experimental data, dashed lines ---
model calculations)
\end{figure}

Satisfactory agreement between available data and theoretical result
(rescaled variant of the model with Skyrme constant $e=3.0$) takes place in
the case of isotopes with isospin $1/2$ and $9/2$ as well, see Table 9.
The numbers with $^\dagger$ in the 3-d
and 5-th column are results of extrapolation made in [\cite{29}].

The nucleus $^{17}{\rm Be}$, similar to $^{15}{\rm Be}$,  has not been observed yet, so,
the b.e. $63.4$~MeV which follows from Table 9 also is
prediction of the chiral soliton approach. This isotope is not stable since
the b.e. of $^{14}{\rm Be}$, $70$~MeV, is greater. For the isotope
$^{19}{\rm B}$ the prediction of our approach is $88.7$~MeV, in good agreement with
the extrapolation result $90.1$~MeV shown in Table 9. Satisfactory
agreement between our prediction and extrapolation of paper [\cite{29}] takes
place for isotopes $^{21}{\rm C}$, $^{23}{\rm N}$, and $^{25}{\rm O}$.

\begin{center}
\begin{tabular}{|l|l|l|l|l|l|l|}
\hline
$A$&$\epsilon_{1/2}^{\rm exp}$&$\epsilon_{9/2}^{\rm exp}$&$\Delta\epsilon_{\rm C}$&
$\Delta\epsilon_{1/2,9/2}^{\rm exp}$
&$\Delta\epsilon_{1/2,9/2}^{\rm th}$&$\Delta\epsilon_{1/2,9/2}^{\rm th*}$\\ \hline
$\rm^{17}O $$-$$\rm^{17}Be$ & $131.8$ & \ ---  & $8.9 $ & \ ---  & $181.3$ & $77.3$ \\ \hline
$\rm^{19}F $$-$$\rm^{19}B $ & $147.8$ & $\;90.1^\dagger$& $10.4$ &$68.1^\dagger$& $162.4$ & $69.5$ \\ \hline
$\rm^{21}Ne$$-$$\rm^{21}C $ & $167.4$ & $118.8^\dagger$ & $11.9$ &$60.5^\dagger$& $148.3$ & $63.5$ \\ \hline
$\rm^{23}Na$$-$$\rm^{23}N $ & $186.6$ & $142.2^\dagger$ & $13.3$ &$57.7^\dagger$& $136.4$ & $58.5$ \\ \hline
$\rm^{25}Mg$$-$$\rm^{25}O $ & $205.6$ & $168.4^\dagger$ & $14.7$ &$51.9^\dagger$& $126.8$ & $54.4$ \\ \hline
$\rm^{27}Al$$-$$\rm^{27}F $ & $224.9$ & $185.8$ & $16.0$ &$55.1$& $118.1$ & $50.7$ \\ \hline
$\rm^{29}Si$$-$$\rm^{29}Ne$ & $245.0$ & $208.2$ & $17.4$ &$54.2$& $110.7$ & $47.6$ \\ \hline
$\rm^{31}P $$-$$\rm^{31}Na$ & $262.9$ & $228.9$ & $18.7$ &$52.7$& $104.2$ & $44.8$ \\ \hline
\end{tabular}
\end{center}
{\bf Table 9.} The binding energies differences for isotopes with isospin
$I=1/2$ (i.e. $N=Z+1$) and $9/2$ ($N=Z+9$) for the original variant, $e=4.12$,
and for the variant with  rescaled constant, $e=3.0$ \\

The results for odd atomic numbers (half integer isospins,
Tables 7$-$9) are presented also in Fig.\ 2. Similarity to behavior
of data and calculation results shown in Fig.\ 1 is evident.

In Table 10 we present also predictions of the chiral soliton approach for
nuclides with $N=Z+11$, or isospin $I=11/2$ for the SK4 and SK6 variants of
the model.
\begin{center}
\begin{tabular}{|l|l|l|l|l|l|l|}
\hline
$A$&$16/\Theta_I^{\rm SK4}$&$\Delta\epsilon_{\rm C}$&
$\epsilon_{11/2}^{\rm SK4}$&$\epsilon_{11/2}^{\rm SK6}$&$\epsilon_{11/2}^\dagger$ \\ \hline
$\rm^{17}Li$ & $\;103.2 $  & $\;7.2$ & $\;28$& $\;23$ &\ ---  \\ \hline
$\rm^{19}Be$ & $\;\;92.5$  & $\;8.7$ & $\;60$& $\;56$ &\ ---  \\ \hline
$\rm^{21}B $ & $\;\;84.7$  & $10.2$  & $\;88$& $\;83$ &\ ---  \\ \hline
$\rm^{23}C $ & $\;\;78.0$  & $11.7$  & $117$ & $113 $ &\ ---  \\ \hline
$\rm^{25}N $ & $\;\;72.4$  & $13.1$  & $143$ & $139 $ &\ ---  \\ \hline
$\rm^{27}O $ & $\;\;67.5$  & $14.4$  & $170$ & $166 $ &\ ---  \\ \hline
$\rm^{29}F $ & $\;\;63.5$  & $15.8$  & $194$ & $191 $ &$187$  \\ \hline
$\rm^{31}Ne$ & $\;\;59.7$  & $17.1$  & $219$ & $216 $ &$212$  \\ \hline
\end{tabular}
\end{center}
{\bf Table 10.} Predictions for the binding energies of nuclides with
$I=11/2$ or $N=Z+11$ calculated from nuclides with $I=3/2$ ($N=Z+3$) \\

For the nuclides $^{29}{\rm F}$ and $^{31}{\rm Ne}$ we obtain fair agreement with the
extrapolation result obtained in [\cite{29}], and the SK6 variant prediction
is more close to it than the SK4 one. Experimental data for this case
and $A\leq 32$ are still lacking.
\section{6. Conclusions}
The estimates of the mass splittings of nuclides with baryon numbers
up to $32$ are made using the properties of the bound states of skyrmions
obtained in the rational map approximation.
The results are in impressive agreement with data, provided the rescaling
of the Skyrme constant $e$ is made --- decrease by about $25$$-$$30\%$ --- which
effectively takes into account the role of nonzero modes, vibration and
breathing. This nonzero modes motion brings the rational map skyrmions,
classical configurations of the one-shell type, in better agreement with
observed nuclear matter distribution: it increases the dimensions of
configuration and, therefore, the moments of inertia. Without rescaling of the
constant $e$ the mass splittings of light nuclear isotopes are overestimated
more than twice. It is known that calculations of nonzero modes contribution
to the energy (mass) of multiskyrmion are closely related to calculations of
the so called Casimir energy and/or loop corrections to the soliton mass
[\cite{31}] which are of the order of $N_c^0 \sim 1$. The estimate of this
contribution was made for the $B=1$ hedgehog-type skyrmion [\cite{31}].
Anyway, these unknown contributions cancel out in the differences of energies
of quantized states we have considered here.

Since 1983 [\cite{2}] it was believed that the chiral soliton approach allows
to describe the properties of baryons (the $B=1$ sector) with accuracy
$\sim 30\%$ and better. We have shown, that quite good description of the binding energy
differences can be obtained also for light nuclei $B=A\leq 32$, when appropriate
rescaling of the
constant $e$ is made, providing the ``nuclear'' variant of the model.
Taking into account the former results on good qualitative description of
binding energies of light hypernuclei [\cite{13}], we conclude
that effective field theories, including the chiral soliton approach, provide
good description of certain basic properties of relatively light nuclei.

We can conclude, that besides the $B=1$ version of the model, one can consider
the ``nuclear'' version of the model with rescaled value of $e$ which is good
for the description of nuclear properties for $A < 30$.
Observation of the predicted states can provide additional support of
the validity of this approach which turned out to be very effective in the
understanding of the properties of mesons and baryons. Studies of other properties
of nuclei (e.g. magnetic moments, radii of different distributions, etc.) is now
one of actual problems.

No stable multineutron nuclides are found within the chiral soliton
approach;  all states  are several tens of MeV above the threshold
and can decay strongly into final states consisting of nucleons.
The states with isospin $T > B/2$ also can be obtained. They can be
interpreted as states consisting of nucleons and $\Delta^-$ isobars,
and some of them can be negative --- as $n...n\Delta^-$. Estimates show,
however, that their energies are higher than thresholds for strong decays.

One of the evident drawbacks of this approach is that one cannot describe
the variations of the isotopical content of nuclides
in space: in the quantization procedure solitons are rotated in the isospace
as a whole. In particular, the description of neutron halo on the periphery of
neutron rich nuclides, established experimentally (see, e.g. [\cite{20}]), is
behind the possibilities of our approach in its present form.
The studies of nonzero modes quantum fluctuations are necessary
for this, since the approach we used corresponds to excitation only of collective
rotational modes.

Careful study of the results presented in Tables 2$-$5
and 7$-$9 allows to reveal the following general feature: for relatively
light nuclei, $A < 20$$-$$22$, theoretical predictions for the difference of
binding energies of isotopes are greater than the experimentally observed 
value
s
of this difference. For $A > 22$$-$$24$ theoretical predictions for this
difference are smaller than data, as a rule. One could reach better agreement
with data by decreasing the value of the constant $e$ at smaller $A$ (thus
providing the additional increase of dimensions of multiskyrmions), and
increasing slightly $e$ at larger $A$. It means, that the relative role
of nonzero modes in formation of the size of nuclei slightly decreases with
increasing $A$, probably, because for large $A$ the size of multiskyrmion is
large enough. This observation shows that further more detailed studies and
refinements would be useful and of interest, although they will not change our
main conclusions concerning the validity of the chiral soliton approach for
description of the binding energies of nuclides.

We are indebted to E.~Konobeevsky, A.~Reshetin, and Yu.~Ryabov for discussions on
the initial stages of the work, stimulating our interest to the problem
considered here. The work is supported by RFBR, grant no. 01-02-16615.

\end{document}